# Bio-Organic Materials Based Resistive Switching Memories

Rahul Deb[1], Debajyoti Bhattacharjee[1] and Syed Arshad Hussain[1]*

[1]Thin Film and Nanoscience Laboratory, Department of Physics, Tripura University- 799022, India

*Corresponding author, Email: sahussain@tripurauniv.ac.in

## Abstract

Resistive switching (RS) devices, based on soft materials such as organic, biomolecules as well as natural plant extracts etc., has emerged as a promising alternative to the conventional memory technologies. They offer simple device structures, low power requirements, rapid switching and compatibility with high-density device integration. Over the last two decades, these classes of materials have been explored for both non-volatile memory and artificial synapse functions. This chapter provides a brief overview of RS fundamentals, their major classifications, key applications, and recent trends in the use of organic and bio-derived materials.

## 1. Introduction

The rapid developments in modern technologies have created an ever-growing demand for ultrahigh-density memory devices with lower power consumption, faster response and reduced fabrication complexity. Numerous limitations of conventional silicon-based memory technologies including high fabrication cost, limited scalability and reusability, have been the key drivers in the search for alternative technologies capable of overcoming these limits. In this context, resistive switching (RS) devices have gained considerable attention due to their inherent ability to overcome these limitations [1]. This trend is clearly observed from the increasing number of publications over the last two decades (figure 1a). RS devices generally consist of a metal/insulator/metal (MIM) structure. The metal contacts act as the electrodes, whereas the middle layer acts as the active component which undergoes a transition between a high resistance state (HRS) and a low resistance state (LRS) upon application of bias voltage on the electrodes. These two states can be used to encode binary information (figure 1b). Depending on the nature of the material and the direction of voltage sweeping, RS devices may exhibit nonvolatile or volatile memory behaviour. Nonvolatile devices retain the LRS or HRS even after removal of the external bias, whereas volatile devices return to the HRS at low voltage [2]. Recent developments in RS materials show a strong shift toward organic, biomolecular and plant-derived systems.

## 2. Fundamentals of Resistive Switching

### 2.1. Working Principle

RS devices exploit the reversible transition between two or more stable resistance states of the active layer material. The fundamental principle of these devices is the modulation of charge transport pathways within the active layer by means of suitable mechanisms such as redox reactions, charge trapping/detrapping, space-charge-limited conduction (SCLC) and conductive filament formation etc. [2]. When the applied bias reaches the set voltage ($V_{SET}$), the material undergoes a transition from HRS to LRS. Depending on the material used, the LRS may persist even after the bias is withdrawn resulting in non-volatile memory. On the application of another suitable voltage sweep, the device may undergo a transition from the LRS to the HRS at the reset voltage ($V_{RESET}$) [3]. The charge transport mechanism in RS devices can be investigated using log-log I-V plots. The slope values from this plot can be used to identify bias regions where the current flow is governed by thermionic emission and Space Charge Limited Conduction (SCLC) etc. [2]. Other suitable fitting schemes can also be utilized in order to verify several conduction models including Schottky Emission, Fouler-Northeim Tunnelling, Trap Assisted Tunneling and Poole-Frenkel mechanism etc. [1, 2].

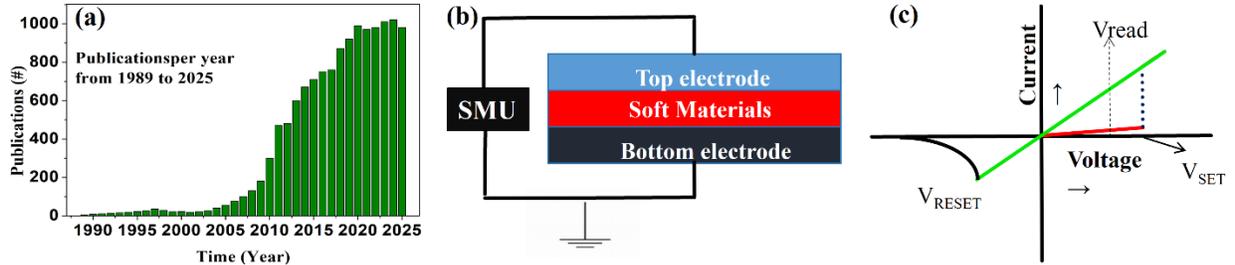

Figure 1. (a) Publications per year from 1989 to 2025. Data collected from web of science (https://apps.webofknowledge.com/) using the search topic 'Resistive Switching'. (b) MIM device structure of RS device. Schematic of I-V response of RS device showing RRAM behaviour.

## 2.2. Memory Logic

Memory implementation in RS devices is based on the encoding of binary data within distinct resistance states. The HRS and LRS correspond to logical "0" and "1," or vice versa, depending on the read voltage logic. The device state (whether "0" or "1") is read by applying a read voltage lower then than $V_{SET}$ or $V_{RESET}$ to ensure noninvasive detection of logic states. Moreover, choosing a lower read voltage also ensure lower power consumption in data retrieval process.

## 2.3. Classification of RS Devices

Based on I-V characteristics, RS devices are broadly classified into write-once-read-many (WORM), resistive-random-access-memory (RRAM), threshold Switching (TS) and complementary resistive switching (CRS) memory [2]. On application of suitable bias voltage, WORM devices exhibit an irreversible transition from HRS to LRS which is ideal for permanent data storage. RRAM devices exhibit reversible transition from HRS to LRS and vice versa depending on applied bias. This type of devices is ideal for rewritable data storage (such as flash drives) as well as processing hardware. TS devices possess only one stable resistance state in unbiased condition. Unlike the other RS device, in TS devices, upon the removal of the switching bias ($V_{SET}$), they return to their original HRS state. CRS devices use complementary resistance states to mitigate sneak-path currents [4].

## 2.4. Performance Parameters

The performance of an RS device is evaluated through several key memory parameters that collectively determine its suitability for practical applications [1]. Table 1 gives an overview of these parameters.

Table 1. Performance Parameters in RS devices and their significance

| Memory Parameter | Description | Significance |
|---|---|---|
| Compliance Current | Maximum allowed current during SET process | Prevents device breakdown and controls filament formation |

| ON/OFF Ratio (Memory Window) | Ratio of LRS current to HRS current at read voltage | Measures readability, noise immunity, and data reliability |
|---|---|---|
| Retention Time | Duration for which LRS/HRS is maintained without bias | Indicates long-term data stability |
| Read Endurance | Number of times the LRS/HRS states can be read by applying read voltage | Indicates how many times the stored data can be read from the device |
| Cycling Stability | Number of successful SET/RESET cycles | Reflects device durability and practical usability |
| Response Speed | Time required for switching events | Critical for high-speed memory applications |
| Device Yield | Percentage of working devices fabricated | Indicates process reliability and materials uniformity |

## 5. Importance of Organic and Biomolecules

Organic and biomolecular materials as well as natural plant extracts have gained significant importance in resistive switching, because they offer a unique combination of structural tunability, low-cost fabrication and compatibility with sustainable electronics. For instance, the molecular frameworks of organic molecules can be systematically modified to adjust donor-acceptor strength, π-conjugation length and functional groups, enabling precise control over the HOMO-LUMO energy gap and the corresponding charge-transport behaviour [1-8]. The ease of processing is another major advantage of these materials. Such materials can be deposited through simple methods under ambient conditions such as drop-casting, spin-coating etc. Moreover, water-soluble and recyclable biomolecules, as well as naturally occurring plant extracts support environmentally friendly device fabrication and open pathways toward transient and degradable electronic systems [2, 8]. On the other hand, Biomolecules and plant extracts introduce additional benefits due to their inherent functional diversity. Their multiple heteroatoms and naturally occurring donor-acceptor units provide intrinsic charge-transfer capabilities, facilitating efficient field-induced conduction. Lastly, the compatibility of organic and bio-derived materials with flexible substrates and their low ecological footprint makes them compelling candidates for future resistive memory, neuromorphic devices and sustainable electronic systems.

## 6. Recent Trends

Organic small molecules, particularly π-conjugated donor-acceptor frameworks such as indolyl, coumarin derivatives, have demonstrated tunable WORM and RRAM characteristics with high memory windows and excellent retention. Here, structural modification at donor sites or incorporation of electron-rich groups significantly alters switching modes (figure 2a) [1]. Performance tuning through hybridization has also emerged as a major trend. For instance, ZnO nanoparticle incorporation into coumarin active layers markedly enhances device yield, endurance, and long-term stability due to oxygen vacancy assisted filament formation combined with organic charge-transfer pathways (figure 2d) [5]. A similar strategy is evident in clay intercalated plant-extract devices, where synthetic Laponite improves retention to 10 years and enables transition from WORM to RRAM, highlighting the role of inorganic trap states in stabilizing switching events (figre 2b) [6]. In parallel, plant derived materials have become prominent due to their biodegradability, natural donor/acceptor groups, and intrinsic oxygen-defect. Devices based on Ipomoea carnea and

Nymphaea nouchali show stable WORM, rewritable read-only, and even neuromorphic synaptic behaviors with long-term physical stability exceeding 360 days. [7]. Furthermore, protein-based systems represent an advancing frontier, offering inherent bio-compatibility, tunable functional groups, and potential for multilevel states, reaffirming the growing momentum toward sustainable and flexible bio-integrated memory technologies [2]. A recent study on the RS behaviour of Lysozyme protein showed that biomaterial-based RS devices can have stability in excess of 10 years (figure 2c) [8].

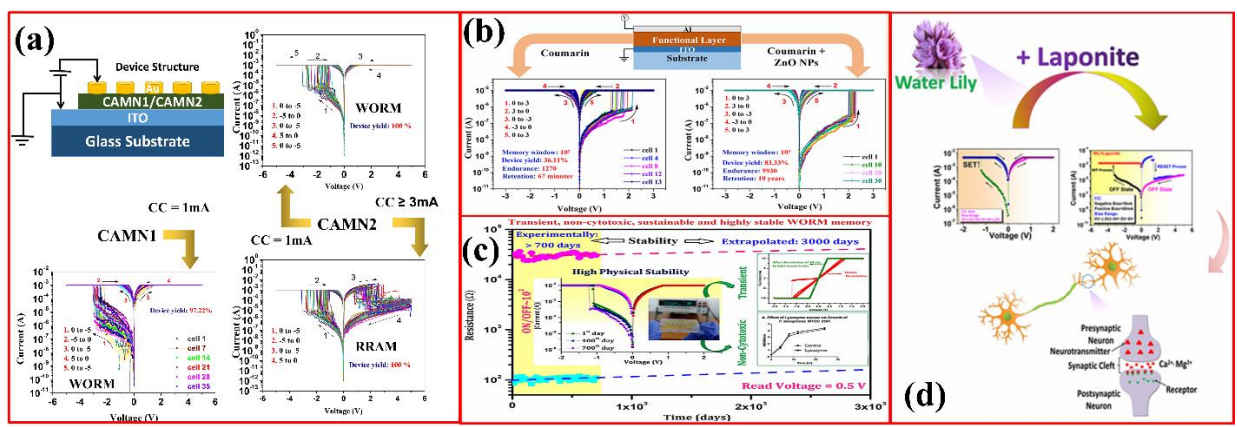

Figure 2. (a) WORM and RRAM type RS in coumarin based devices modulated by the presence and absence of strong donor (-OCH$_3$) group [1]; (b) ZnO induced performance enhancement of coumarin based RS device [5]; (c) Lysozyme protein-based RS device showing physical stability of 10 years; (d) Artificial synapse behaviour in laponite incorporated Water Lily based RS device [6].

## 7. Applications

RS devices have attracted considerable interest for a broad range of electronic applications. One of the most prominent uses is in nonvolatile memory storage. Here both WORM and RRAM devices may serve as alternatives to conventional flash memory. WORM RS devices are particularly suitable for secure data archiving and tamper-proof information storage, whereas, RRAM devise may serve as replacement for silicon based editable storage devices. The capability of resistive devices to operate within compact crossbar arrays facilitates large-scale memory integration. The analog RS behaviour of certain organic and hybrid materials makes them ideal for building artificial synapses for brain-inspired computing. Collectively, these diverse applications demonstrate the versatility and technological potential of RS devices in next-generation electronics.

## 8. Challenges and Future Prospects

Despite promising progress, several challenges remain before resistive switching devices can be widely adopted. Variability in switching parameters such as threshold voltages, endurance, and memory windows-remains a major concern. Device-to-device reproducibility, especially in solution-processed films, requires improved control over thickness, morphology, and molecular orientation. Understanding the interplay between molecular structure, charge transport, and switching pathways is necessary to reduce stochastic behaviour. Scalability in crossbar arrays requires consistent CRS performance. Lastly, a robust understanding of the switching and conduction mechanisms in the RS devices still remains incomplete.

# 9. Conclusion

Resistive switching offers a powerful and versatile route for next-generation memory and electronic devices. Herein, organic small molecules have emerged as promising class of materials due to their tunable electronic properties, sustainability, and ease of processing. Recent advancements demonstrate significant improvements in memory windows, switching voltages, and device reliability. Continued research into material design, mechanism understanding, and device optimization will pave the way for practical deployment in high-density memory, neuromorphic computing and flexible electronics.

## Acknowledgement

S.A.H. is grateful to CSIR (No. 03/1504/23/EMR-II) for the financial support to carry out this research work. Mr. Rahul Deb is thankful CSIR for financial assistance via CSIR-SRF (Direct) award vide File. No. 09/0714(23956)/2025-EMR-I to carry out this research work